\documentclass[aps,prl,twocolumn,longbibliography] {revtex4-1}
\usepackage{amsmath,graphicx,amsfonts}

\usepackage{times}
\usepackage[colorlinks=true,linkcolor=blue,urlcolor=blue,citecolor=blue, breaklinks=true,hypertexnames=false]{hyperref}

\usepackage{amssymb}
\usepackage{bbm}

\usepackage[utf8]{inputenc}
\DeclareUnicodeCharacter{03BC}{\mbox{$\mu$}}

\DeclareUnicodeCharacter{03B4}{$\delta$}
\DeclareUnicodeCharacter{2009}{-}

\usepackage[english]{babel}
\addto\captionsenglish{}

\makeatletter
\def\bbl@set@language#1{%
	\edef\languagename{%
		\ifnum\escapechar=\expandafter`\string#1\@empty
		\else\string#1\@empty\fi}%
	\@ifundefined{babel@language@alias@\languagename}{}{%
		\edef\languagename{\@nameuse{babel@language@alias@\languagename}}%
	}%
	\select@language{\languagename}%
	\expandafter\ifx\csname date\languagename\endcsname\relax\else
	\if@filesw
	\protected@write\@auxout{}{\string\select@language{\languagename}}%
	\bbl@for\bbl@tempa\BabelContentsFiles{%
		\addtocontents{\bbl@tempa}{\xstring\select@language{\languagename}}}%
	\bbl@usehooks{write}{}%
	\fi
	\fi}
\newcommand{\DeclareLanguageAlias}[2]{%
	\global\@namedef{babel@language@alias@#1}{#2}%
}
\makeatother

\DeclareLanguageAlias{en}{english}

\makeatletter
\def\@bibdataout@aps{%
	\immediate\write\@bibdataout{%
		@CONTROL{%
			apsrev41Control%
			\longbibliography@sw{%
				,author="08",editor="1",pages="1",title="0",year="1"%
			}{%
				,author="08",editor="1",pages="1",title="",year="1"%
			}%
		}%
	}%
	\if@filesw \immediate \write \@auxout {\string \citation {apsrev41Control}}\fi 
}
\makeatother

\def\B {\scriptscriptstyle {B}}
\def\Q {\scriptscriptstyle {Q}}

\date\today
\begin{document}
	
\title{Mediated interaction between polarons  in a one-dimensional Bose gas}
\author{Aleksandra Petkovi\'{c} and Zoran Ristivojevic}
\affiliation{Laboratoire de Physique Th\'{e}orique, Universit\'{e} de Toulouse, CNRS, UPS, 31062 Toulouse, France}
	
\begin{abstract}
We study a weakly-interacting one-dimensional Bose gas with two impurities coupled locally to the boson density. We derive analytical results for the induced interaction between the impurities at arbitrary coupling and separation $r$. At $r\lesssim \xi$, where $\xi$ denotes the healing length of the Bose gas, the interaction is well described by the mean-field contribution. Its form changes as the coupling is increased, approaching a linear function of $r$ at short distances in the regime of strong coupling. The mean-field contribution decays exponentially at arbitrary coupling for $r\gg\xi$. At such long distances, however, the effect of quantum fluctuations becomes important, giving rise to a long-ranged quantum contribution to the induced interaction. At longest distances it behaves as $1/r^3$, while at strong coupling we find an intermediate distance regime with a slower decay, $1/r$. The quantum contribution in the crossover regime is also calculated. The induced interaction between impurities (i.e., polarons) is attractive and leads to the formation of their bound state, known as bipolaron. We discuss its binding energy.
\end{abstract}
\maketitle
	
The concept of mediated interaction plays a pivotal role in physics. Within the standard model, the fundamental interactions between matter particles are mediated by bosonic fields \cite{Halzen}. In quantum electrodynamics, the Casimir effect denotes the interaction between metallic plates mediated by the virtual excitations in the vacuum \cite{kardar_friction_1999}. In condensed matter, the formation of Cooper pairs in conventional Bardeen-Cooper-Schrieffer superconductors  occurs due to the attraction between electrons mediated by the quanta of lattice vibrations \cite{Tinkham}. Another example is the Ruderman–Kittel–Kasuya–Yosida exchange interaction between nuclear magnetic moments or localized electrons mediated by the conduction electrons in metals \cite{Jensen}. Recently, related phenomena have been experimentally studied with ultra-cold gases, which give rise to an attractive interaction between foreign particles -- impurities \cite{cetina_ultrafast_2016,desalvo_observation_2019,edri_observation_2020}.

A mobile impurity interacting with a bath of quantum particles transforms into a polaronic quasiparticle with distinct features from the original particle \cite{mahan}. First obtained for electrons in ionic crystals \cite{landau_effective_1948}, the latter scenario also applies for impurities in ultra-cold Bose gases. Many theoretical \cite{astrakharchik_motion_2004,cucchietti_strong-coupling_2006,casteels_many-polaron_2011,rath_field-theoretical_2013,shashi_radio-frequency_2014,li_variational_2014,levinsen_impurity_2015,ardila_impurity_2015,christensen_quasiparticle_2015,camacho-guardian_landau_2018,naidon_two_2018,drescher_real-space_2019,levinsen_quantum_2021} and experimental \cite{catani_quantum_2012,spethmann_dynamics_2012,hu_bose_2016,jorgensen_observation_2016,yan_bose_2020,skou_non-equilibrium_2021} papers considered the latter system.

Studies of impurities in one-dimensional Bose gases are of particular interest since reduced dimensionality enhances the role of quantum fluctuations, leading to phenomena where the mean-field description is insufficient  \cite{pitaevskii_bose-einstein_2003}. In Bose gas environments various properties of a single polaron have been studied \cite{fuchs_spin_2005,sacha_self-localized_2006,zvonarev_spin_2007,schecter_dynamics_2012,petkovic_dynamics_2016,grusdt_bose_2017,parisi_quantum_2017,volosniev_analytical_2017,kain_analytical_2018,panochko_mean-field_2019,jager_strong-coupling_2020}. In cases when two (or more) impurities are present in the system, the induced interaction between them due to the interaction with particles of the medium is one of the most basic problems \cite{recati_casimir_2005,schecter_phonon-mediated_2014,reichert_casimir-like_2019,reichert_fluctuation-induced_2019,brauneis_impurities_2021}. For identical impurities that are locally and weakly coupled to the Bose gas, an exponentially small attractive interaction proportional to $e^{-2r/\xi}$ was found in Ref.~\cite{recati_casimir_2005}. Here $r$ denotes the impurity separation, while $\xi$ is the healing length of the Bose gas. However, the effect of quantum fluctuations gives rise to another contribution to the induced interaction of a long-range nature \cite{schecter_phonon-mediated_2014}. It behaves as $\xi^3/r^3$ at $r\gg \xi$ and therefore becomes dominant at long distances \cite{schecter_phonon-mediated_2014,reichert_casimir-like_2019}.

\begin{figure}
\includegraphics[width=0.8\columnwidth]{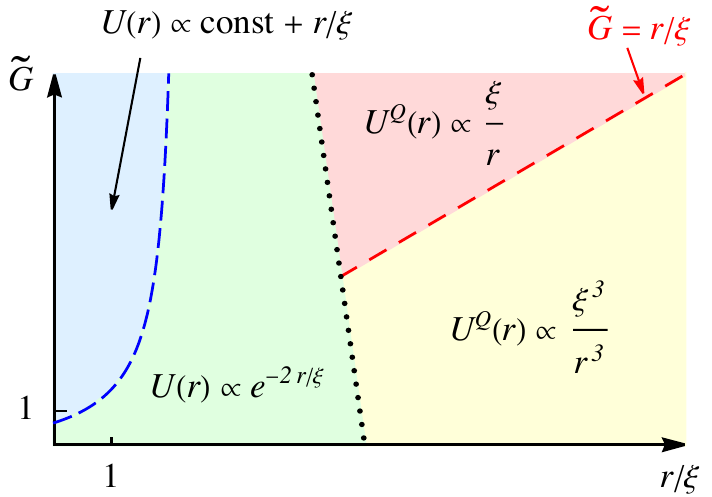}
\caption{Schematic diagram of different regimes of the induced interaction between two polarons in a weakly-interacting one-dimensional Bose gas. At short separations $r$, the mean-field interaction $U(r)$ is dominant, while for $r$ longer than a few $\xi$, the long-range interaction $U^{\Q}(r)$ that originates from  quantum fluctuations prevails. The two qualitatively different contributions to the induced interaction are separated by the dotted line, which depends weakly on $\widetilde{G}$ and $\gamma$. At $\gamma=0.1$, its position is near $5.4r/\xi$ at $\widetilde{G}\ll 1$ and $3.6 r/\xi$ at  $\widetilde{G}\gg 1$.} \label{fig1}
\end{figure}

In this paper we take advantage of weak repulsion between bosons to calculate analytically the induced interaction between impurities at \emph{arbitrary} coupling. This is possible due to the existence of an analytical solution of the corresponding Gross-Pitaevskii equation, enabling us to find exactly the mean-field contribution to the induced interaction, which is dominant at $r\lesssim \xi$. In the complementary regime $r\gg\xi$, we apply the scattering approach to find explicit results for the quantum contribution to the induced interaction. It is long-ranged and shows two characteristic regimes, see Fig.~\ref{fig1}. Our theory has direct implication for the many-body physics with polarons, which will exhibit clusterization into multi-polaronic bound states due to the induced attractive interaction.

We study a one-dimensional system of weakly-interacting bosons with repulsive short-range interaction of the strength $g$. Consider two impurities at separation $r$, locally coupled to the Bose gas density, which is modeled by the potential $V(x)=G\left[\delta(x+r/2)+\delta(x-r/2)\right]$. At the mean-field level, the system can be described by the Gross-Pitaevskii equation \cite{pitaevskii_bose-einstein_2003}
\begin{gather}\label{eq:GPE}
i\hbar\frac{\partial \psi_0(x,t)}{\partial t}=\left[-\frac{\hbar^2}{2m}\frac{\partial^2 }{\partial x^2}+g|\psi_0(x,t)|^2+V(x)\right] \psi_0(x,t).
\end{gather}
Here  $m$ denotes the mass of bosons. For convenience, we study the system with periodic boundary conditions. 

Let us first consider heavy, static impurities and at a later stage account for their kinetic energy using perturbation theory. In this case we can assume the solution of Eq.~(\ref{eq:GPE}) in the form $\psi_0(x,t)=\psi_0(x)e^{-i \mu t/\hbar}$, where $\mu$ denotes the chemical potential. Equation (\ref{eq:GPE}) then reduces to the nonlinear eigenvalue problem for $\psi_0(x)$, which has many solutions. Among them, we seek for the one with the smallest energy. The corresponding eigenfunction $\psi_0(x)$ is nodeless at finite coupling $G$. However, in the limit $G\to+\infty$, the boson density must be completely depleted at the impurity positions, leading to two nodes in $\psi_0(x)$. The ground-state energy of the system in the mean-field approximation  is given by
\begin{align}\label{eq:E0def}
E_0(r)=\mu n L-\frac{g}{2}\int_{-L/2}^{L/2} dx |\psi_0(x)|^4.
\end{align}
The chemical potential $\mu$ entering $E_0(r)$ in Eq.~(\ref{eq:E0def}) should be eventually expressed in terms of the boson density $n$ from the normalization condition 
\begin{align}\label{eq:n}
n=\frac{1}{L}\int_{-L/2}^{L/2} dx |\psi_0(x)|^2.
\end{align}
By $L$ is denoted the system size. The mean-field contribution to the induced interaction between the impurities mediated by the Bose gas is defined by $U(r)=E_0(r)-E_0(r\to\infty)$. It vanishes at $r\to\infty$.

\begin{figure}
\includegraphics[width=\columnwidth]{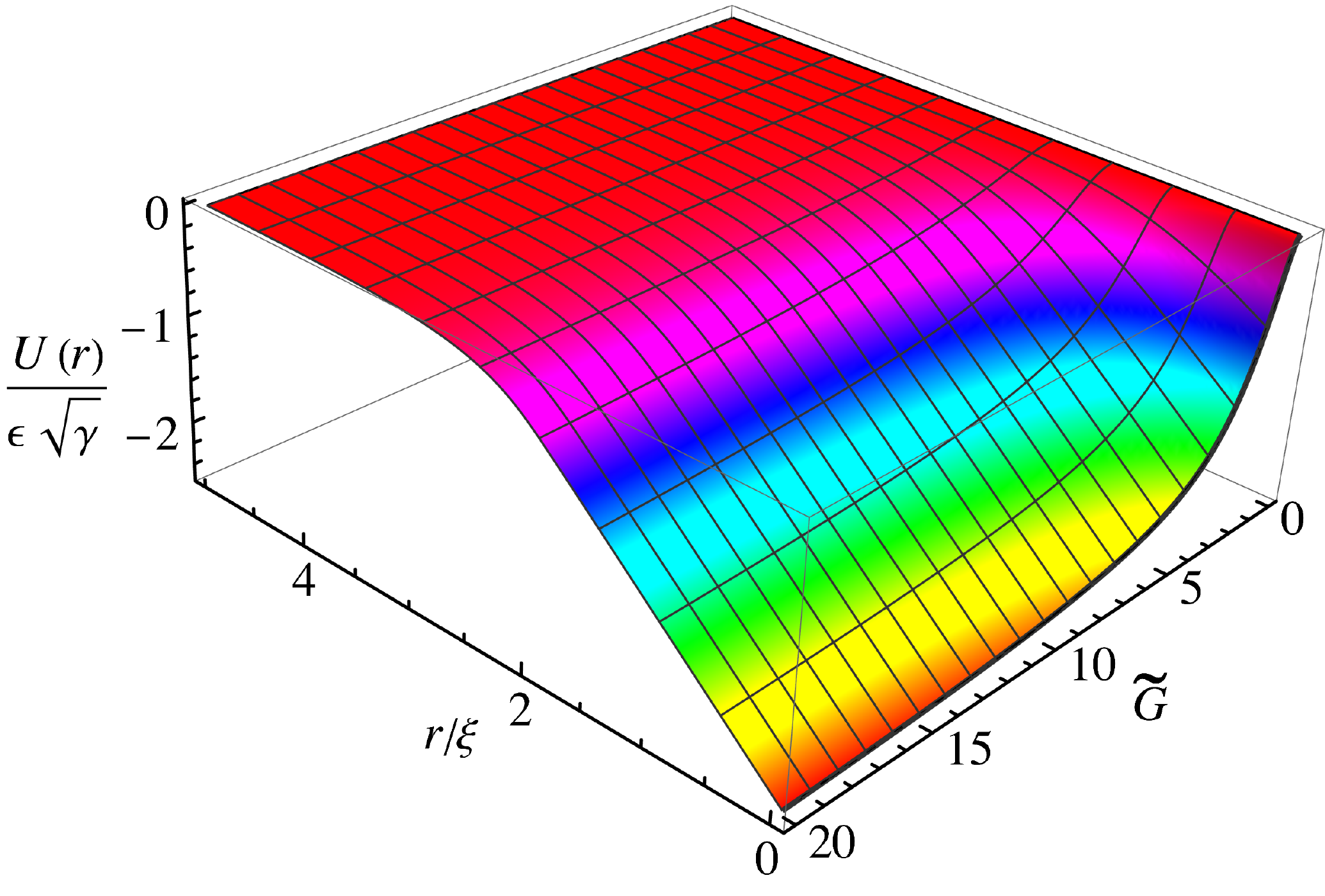}
\caption{Plot of the mean-field interaction $U(r)$ given by Eq.~(\ref{eq:E0}) as a function of the dimensionless distance between impurities $r/\xi$ and the dimensionless coupling $\widetilde{G}=G\sqrt{\gamma}/g$. The interaction (\ref{eq:E0}) is bounded, $-8\epsilon\sqrt{\gamma}/3\le U(r)\le 0$; colors on the plot correspond to its value.} \label{fig2}
\end{figure}

At $G=0$, corresponding to the absence of impurities, we find $\psi_0(x)=\sqrt{\mu/g}$. The condition (\ref{eq:n}) 
then leads to the chemical potential of the weakly-interacting Bose gas, $\mu_0=g n$. At $G>0$, the local density near the impurity positions deviates from the constant value in spatial regions on the order of the healing length of the Bose gas, $\xi=\hbar/\sqrt{m\mu_0}$. Up to the phase factor, for the family of solutions of Eq.~(\ref{eq:GPE}) we find \cite{reichert_casimir-like_2019,Note1}
\footnotetext[1]{In the solution (\ref{eq:psi0}) and Eqs.~(\ref{eq:bc}) we replaced $\mu$ by $\mu_0$, i.e., $\xi_\mu=\hbar/\sqrt{m\mu}$ by $\xi=\hbar/\sqrt{m\mu_0}$, since we study the system at the leading order in weak interaction between bosons.}
\begin{align}\label{eq:psi0}
\psi_0(x)=\sqrt\frac{\mu}{g}\times \begin{cases}
\sqrt{\frac{2a}{1+a}}\,\textrm{cd}\left(\sqrt\frac{2}{1+a}\frac{x}{\xi};a\right), &|x|<\frac{r}{2},\\
\tanh\left(\frac{|x|}{\xi}-\frac{r}{2\xi}+b\right), &|x|>\frac{r}{2}.
\end{cases}
\end{align}
Here $\textrm{cd}(x;a)$ is the Jacobi elliptic function. We consider the case of identical impurities  and therefore the solution (\ref{eq:psi0}) is an even function, $\psi_0(x)=\psi_0(-x)$. The parameters $0\le a\le1$ and $b$ should be obtained from the continuity of $\psi_0(x)$ at the impurity position $x=r/2$ and the jump in the derivative, $\psi_0'(r/2+0)-\psi_0'(r/2-0)=2mG \psi_0(r/2)/\hbar^2$. This yields
\begin{subequations}
\label{eq:bc}	
\begin{gather}\label{eq:bc1}
\tanh(b)=\sqrt{\frac{2a}{1+a}}\,\textrm{cd}\left(\tilde r;a \right),\\
\widetilde{G} = \frac{1-a}{\sqrt{8a(1+a)}}\frac{1}{\textrm{cd}\left(\tilde r;a \right)}\left[\frac{1+\sqrt{a}\,\textrm{sn}(\tilde r;a)} {\textrm{dn}(\tilde r;a)}\right]^2,\label{eq:bc2}
\end{gather}
\end{subequations}
where $\tilde r={r}/{\sqrt{2(1+a)}\xi}$ and $\widetilde{G}={G}/{\xi \mu_0}$ is the dimensionless impurity strength. By $\textrm{sn}(x;a)$ and $\textrm{dn}(x;a)$ are denoted the Jacobi elliptic functions. The induced mean-field contribution to the interaction for the wavefunction (\ref{eq:psi0}) then takes the form
\begin{align}\label{eq:E0}
U(r)={}&2\;\!\epsilon\sqrt{\gamma}\, \biggl\{\frac{4}{3}-\frac{5-3a^2-2a}{3\sqrt{2}(1+a)^{3/2}} \tilde r+\frac{4\sqrt{2}\, \mathcal{E}\left(\tilde r;a\right)}{3\sqrt{1+a}}\notag\\
& -\frac{\sqrt{2a}\, \textrm{cd}(\tilde r;a)}{3(1+a)^{3/2}} \left[\frac{1+\sqrt{a}\,\textrm{sn}(\tilde r;a)}{\textrm{dn}(\tilde r;a)}\right]^2[3+5a\notag\\
&-4\sqrt{a}(1+a)\,\textrm{sn}(\tilde r;a)]\biggr\}-2E^{\B}(\widetilde{G}),
\end{align} 
where we employed Eqs.~(\ref{eq:bc}). In Eq.~(\ref{eq:E0}) we have introduced $\epsilon=\hbar^2 n^2/2m$ and $\gamma=mg/\hbar^2 n\ll 1$ is the dimensionless parameter describing the interaction strength between the particles of the Bose gas. By $\mathcal{E}(x;a)$ is denoted the Jacobi elliptic function. The parameter $a$ depends on the impurity strength $\widetilde{G}$ and separation $r$ through the condition (\ref{eq:bc2}). In Eq.~(\ref{eq:E0}), $E^{\B}(\widetilde{G})$ denotes the difference of the ground-state energy of the system with and without an impurity, which is given by
\begin{align}\label{eq:EB}
E^{\B}(\widetilde{G})={}& \epsilon\sqrt\gamma \left(\frac{8}{3}-2\eta-\frac{2\eta^3}{3}\right),\quad \eta=\frac{2}{\widetilde{G}+\sqrt{4+\widetilde{G}^2}}.
\end{align}
This result also has another interpretation. A single impurity very strongly coupled to the Bose gas leads to a complete depletion of the boson density at its position. The resulting energy increase of the system, which is given by Eq.~(\ref{eq:EB}) taken at $\widetilde{G}\to+\infty$, coincides with the boundary energy of the Bose gas \cite{gaudin_boundary_1971,reichert_exact_2019}. We emphasize that Eqs.~(\ref{eq:E0}) and (\ref{eq:EB}) are exact with respect to the impurity strength $\widetilde{G}$, but they are calculated at the lowest order in  $\gamma$. In Fig.~\ref{fig2} is shown the induced mean-field interaction (\ref{eq:E0}).

Explicit analytical result for $U(r)$ can be obtained once the parameter $a$ is eliminated from Eq.~(\ref{eq:E0}) using the condition (\ref{eq:bc2}). In the regime $r\ll \xi$, calculating $a$ to the linear order in $r/\xi$, we obtain
\begin{align}\label{eq:Ushort}
U(r)=E^{\B}(2\widetilde{G})-2E^{\B}(\widetilde{G})+\frac{4\;\!\epsilon\sqrt{\gamma}\,\widetilde{G}^2 }{\left(\widetilde{G}+\sqrt{1+\widetilde{G}^2}\,\right)^2}\frac{r}{\xi},
\end{align}
where $E^{\B}$ is defined by Eq.~(\ref{eq:EB}). The interaction (\ref{eq:Ushort}) is exact with respect to the impurity strength $\widetilde{G}$. Neglected higher-order terms in Eq.~(\ref{eq:Ushort}) are small for $r\ll \xi$ at any $\widetilde{G}$. However, they are small even for $r=\xi$ at $\widetilde{G}\gg 1$, tending to zero at least as $1/\widetilde{G}^2$. In the latter regime we can obtain $U(r)$ differently. Performing the expansion of Eq.~(\ref{eq:bc2}) at $a\ll 1$, we obtain $\widetilde{G}\simeq 1/\sqrt{8a}\cos(r/\sqrt{2}\xi)$. Requiring this expression to be positive, we obtain the condition on the distances, $r<\pi \xi/\sqrt{2}$, while the corresponding interaction at $\widetilde{G}\to+\infty$ is given by the exact expression $U(r)=-\epsilon\sqrt{\gamma}\left(8/3-r/\xi\right)$. This follows by setting $a=0$ in Eq.~(\ref{eq:E0}) or, equivalently, from the result (\ref{eq:Ushort}) at $\widetilde{G}\to+\infty$. Therefore, the interaction (\ref{eq:E0}) is accurately described by the linear function (\ref{eq:Ushort}) at impurity separations that can be even longer than $\xi$ in the regime of strong coupling, $\widetilde{G}\gg 1$.

Linear form of the  interaction (\ref{eq:Ushort}) at $\widetilde{G}\gg 1$ can be understood using a simplified approach where we approximate the wavefunction (\ref{eq:psi0}) as $\psi_0(x)=\sqrt{\mu/g}$ at $|x|>r/2$ and $\psi_0=0$ otherwise. This can be justified by noting that $|\psi_0(x)|^2$ is a convex, even function for $|x|<r/2$ and thus it satisfies $|\psi_0(x)|^2\le |\psi_0(0)^2|\sim na\sim n /\widetilde{G}^2\ll n$ at $r<\pi\xi/\sqrt{2}$. From Eqs.~(\ref{eq:E0def}) and (\ref{eq:n}) we then easily recover the term $\epsilon\sqrt{\gamma}\;\! r/\xi$ in the interaction, but not the constant term. This must be the case as the latter effect cannot be described by the simplified form for $\psi_0(x)$, since the density depletion of the characteristic size $\xi$ at $|x|>r/2$, near the impurities, is neglected.

At $r\gg \xi$, the interaction (\ref{eq:E0}) can be explicitly evaluated by substituting $r/\xi=\sqrt{2(1+a)}[K(a)-c\,]$ in Eqs.~(\ref{eq:bc2}) and (\ref{eq:E0}). This parametrization is motivated by the identity $\textrm{cd}\boldsymbol{(}K(a);a\boldsymbol{)}=0$, where $K(a)$ denotes the complete elliptic integral of the first kind. After performing the expansion around $a=1$, we find $\widetilde{G}=2/\sinh(2c)$, and 
\begin{align}\label{eq:Ularger}
U(r)=-\frac{32\;\!\epsilon\sqrt{\gamma}\,\widetilde{G}^2 }{\left(2+\sqrt{4+\widetilde{G}^2}\,\right)^2} e^{-\frac{2r}{\xi}}.
\end{align} 
The interaction (\ref{eq:Ularger}) is valid at arbitrary $\widetilde{G}$. In the special case of weak coupling, \mbox{$\widetilde{G}\ll1$}, it reduces to the known result \cite{recati_casimir_2005,reichert_casimir-like_2019,reichert_field-theoretical_2019},
$U(r)=-2\epsilon\sqrt\gamma\, \widetilde{G}^2e^{-\frac{2r}{\xi}}$. Unlike Eq.~(\ref{eq:Ularger}), the latter result applies at any $r$. In Fig.~\ref{fig1} are illustrated different regimes of the induced interaction. We notice that the result (\ref{eq:Ularger}) can be understood classically. Each of the impurities produces the disturbance in the boson density of the characteristic size $\xi$. At separations of several $\xi$, the two disturbances practically do not overlap, resulting in an exponentially decaying mean-field interaction.

One should be aware, however, that at distances longer than a few $\xi$ the quantum contribution $U^{\Q}(r)$ to the induced interaction becomes important. It is given by \cite{jaekel_casimir_1991,kenneth_casimir_2008,reichert_fluctuation-induced_2019}
\begin{align}\label{eq:Uscatteringfull}
U^{\Q}(r)=\frac{\hbar v}{2\pi}\,\mathrm{Im}\int_0^\infty dk \ln\left[1-\mathbbm{r}(k)^2 e^{2i k r}\right],
\end{align}
where $v=\hbar n\sqrt\gamma/m$ denotes the sound velocity. The central quantity in Eq.~(\ref{eq:Uscatteringfull}) is $\mathbbm{r}(k)$, which is the reflection amplitude of the Bogoliubov quasiparticle with wavevector $k$ on the potential of the single impurity placed at the origin in the Bose gas. The corresponding scattering problem is obtained by studying the quantum correction, $\hat\psi_1(x,t)$, to the mean-field single-particle bosonic operator, $\psi_0(x,t)$. The former can be understood as a superposition of Bogoliubov quasiparticles. It satisfies the linear equation \cite{pitaevskii_bose-einstein_2003}
\begin{align}\label{eq:psi1}
i\hbar \frac{\partial \hat\psi_1(x,t)}{\partial t}={}&\left[-\frac{\hbar^2 }{2m}\frac{\partial^2}{\partial x^2}+2g |\psi_0(x)|^2+V(x)\right]\hat\psi_1(x,t)\notag\\
&+g\psi_0(x)^2 \hat\psi_1^\dagger(x,t).
\end{align}
Here $V(x)=G\delta(x)$ is the impurity potential, and $\psi_0(x)=\sqrt{n}\tanh\boldsymbol{(}|x|/\xi+\textrm{arctanh}(\eta)\boldsymbol{)}$, which is obtained from the more general solution (\ref{eq:psi0}) replacing $G\to G/2$ and $r\to 0$.

The reflection amplitude for the scattering problem (\ref{eq:psi1}) can be calculated using the standard methods \cite{kovrizhin_exact_2001}. At small momenta and arbitrary $\widetilde{G}$ we obtained
\begin{align}\label{eq:rkzero}
\mathbbm{r}(k)=-\frac{i}{2}\left(\frac{2+\widetilde{G}^2}
{\sqrt{4+\widetilde G^2}}+\widetilde G-1\right)\xi k.
\end{align}
The neglected subleading terms in Eq.~(\ref{eq:rkzero}) are small at $\xi k\ll 1$ in the regime of weak coupling, $\widetilde{G}\ll 1$. However, they can be omitted under the more stringent condition $\xi k\ll 1/\widetilde{G}$ in the regime of strong coupling, $\widetilde{G}\gg 1$. This signals the existence of another regime at intermediate momenta, $1/\widetilde{G}\ll \xi k\ll 1$, in the latter case. 
Therefore, we should study the case $\xi k\sim 1/\widetilde{G}\ll 1$, where we found
\begin{align}\label{eq:rcross}
\mathbbm{r}(k)=\frac{\xi k}{\xi k+i/\widetilde{G}}.
\end{align} 
Equation (\ref{eq:rcross}) describes the crossover between the regimes $1/\widetilde{G}\ll \xi k$ with $\mathbbm{r}(k)=1$, and $\xi k\ll 1/\widetilde{G}$ with $\mathbbm{r}(k)=-i\widetilde{G}\xi  k$, which is a special case of the more general result (\ref{eq:rkzero}).

We are now prepared to evaluate the quantum contribution to the induced interaction between impurities. Substitution of the reflection amplitude (\ref{eq:rkzero}) in Eq.~(\ref{eq:Uscatteringfull}) leads to
\begin{align}\label{eq:Uqr3}
U^{\Q}(r)=-\frac{1}{16\pi}\;\! \epsilon\gamma {\left(\frac{2+\widetilde{G}^2}
	{\sqrt{4+\widetilde G^2}}+\widetilde G-1\right)^2}\, \frac{\xi^3} {r^3}.
\end{align}
The long-range interaction (\ref{eq:Uqr3}) applies at distances $r\gg \xi\, \textrm{max}(1,\widetilde{G})$. In the regime of weak coupling, $\widetilde{G}\ll 1$, it is in agreement with the corresponding result of Refs.~\cite{schecter_phonon-mediated_2014,reichert_casimir-like_2019,reichert_field-theoretical_2019}. Our result (\ref{eq:Uqr3}), however, applies at arbitrary coupling $\widetilde{G}$. It gives explicit dependence of the induced interaction on $\widetilde{G}$.

The interaction (\ref{eq:Uqr3}) becomes inaccurate as the distance decreases toward the crossover regime, $r\sim \xi \widetilde{G} \gg\xi$. There we should use the reflection amplitude (\ref{eq:rcross}), yielding
\begin{align}\label{eq:crossover}
U^{\Q}(r)=\frac{\epsilon\gamma}{2\pi}\frac{\xi}{r} \int_0^{\infty} dz \ln\left(1-\frac{z^2 e^{-z}}{(z+2r/\xi\widetilde{G}\,)^2}\right). 
\end{align} 
Equation (\ref{eq:crossover}) applies at $\widetilde{G}\gg 1$ for arbitrary $r$. In the regime of intermediate distances, $\xi\ll r\ll \xi \widetilde{G}$, from Eq.~(\ref{eq:crossover}) we find the induced long-range interaction,
\begin{align}\label{eq:1/r}
U^{\Q}(r)=-\frac{\pi}{12}\;\!\epsilon\gamma\,\frac{\xi}{r}.
\end{align}
The interaction (\ref{eq:1/r}) shows much slower decay than the result~(\ref{eq:Uqr3}), valid  in the regime of longest distances.  At $r\gg \xi \widetilde{G}$, Eq.~(\ref{eq:crossover}) reduces to Eq.~(\ref{eq:Uqr3}) evaluated at $\widetilde{G}\gg 1$. We notice that the interaction (\ref{eq:1/r}) coincides with a general result for the Casimir interaction of a massless scalar one-dimensional field with two strong $\delta$-function scatterers \cite{milton_casimir_2004}. In Fig.~\ref{fig1} are shown our results for the induced interaction at arbitrary distances and coupling between the impurities and the Bose gas.

The induced mean-field interaction (\ref{eq:E0}) was evaluated for heavy, static impurities. This is not a fundamental limitation of our study since the dynamics of impurities can be accounted for by studying their kinetic energy in perturbation theory. The latter is controlled by small parameter $m/M$, where  $M$ denotes the impurity mass. The correction to the ground-state energy (\ref{eq:E0def}) can be straightforwardly expressed in terms of $\psi_0$ of Eq.~(\ref{eq:psi0}) as 
\begin{align}\label{eq:deltaE}
\Delta E_0(r)=\frac{\hbar^2}{M}\int_{-L/2}^{L/2}dx\, \psi_0\left(-\frac{1}{4}\frac{\partial^2 \psi_0}{\partial x^2}-\frac{\partial^2 \psi_0}{\partial r^2}\right).
\end{align}
Here we assumed that the center of mass of the system is motionless. This is possible since the total momentum commutes with the Hamiltonian and thus it is a conserved quantity, taken to be zero. Equation (\ref{eq:deltaE}) gives a small positive correction to the interaction (\ref{eq:E0}), proportional to  $(m/M)\epsilon\sqrt{\gamma}$.

The induced attraction between impurities mediated by the surrounding Bose gas will favor the formation of their bound state, called bipolaron \cite{roberts_impurity_2009,casteels_bipolarons_2013,camacho-guardian_bipolarons_2018,reichert_field-theoretical_2019,pasek_induced_2019,petkovic_density_2020}. Two heavy impurities at the same position increase the system energy by $E^{\B}(2\widetilde{G})$, while the increase is $2E^{\B}(\widetilde{G})$ when they are far apart. Here $E^{\B}$ is defined by Eq.~(\ref{eq:EB}).  The difference $2E^{\B}(\widetilde{G})-E^{\B}(2\widetilde{G})$ is positive and defines the bipolaron binding energy. A finite impurity mass $M$ gives rise to two effects. Firstly, there is a correction to the induced interaction, as discussed in the previous paragraph. It increases the binding energy on the order of $(m/M)\epsilon\sqrt{\gamma}$. Secondly, at finite $M$ the bound state  should be described quantum-mechanically. It acquires a finite spread around $r=0$, leading to a positive zero-point motion that decreases the binding energy. At $\widetilde{G}\gg 1$, the corresponding energy change is proportional to $(m\sqrt{\gamma}/M)^{1/3} \epsilon\sqrt{\gamma}$. This follows from the study of the Schr\"odinger equation in the linear potential determined by Eq.~(\ref{eq:Ushort}). Here we must note that in order to find the lowest energy state for two massive polarons, it is sufficient to consider the interaction potential at small separations, which is linear. The wavefunction is localized in the region where the potential is linear, rapidly vanishing in the forbidden region. The opposite signs and different parameters in the two corrections to the binding energy leave the possibility to fine tune it varying $\gamma$, which describes the interaction strength of the Bose gas. The detailed study of properties of bipolarons is postponed for a future work.

In conclusion, we have analytically calculated the induced interaction between polarons in a weakly-interacting Bose gas. We have studied the general case of \emph{arbitrary} coupling of impurities to the Bose gas $\widetilde{G}$ and separation $r$, and obtained the results (\ref{eq:Ushort}), (\ref{eq:Ularger}), (\ref{eq:Uqr3}), and  (\ref{eq:1/r}) that apply in four characteristic regions, see  Fig.~\ref{fig1}. The induced interaction is attractive and will lead to the formation of a bound state of polarons. An interesting extension of this paper would be a study of a macroscopic number of polarons in the Bose gas and their clusterization into multi-polaron states. We notice that $N$ heavy Bose polarons in the Bose gas will form a  $N$-polaron bound state since $E^{\B}(N\widetilde{G})< N E^{\B}(\widetilde{G})$. However, a more realistic case should account for the polaron mass and mutual repulsion between them, which will favor the creation of smaller clusters.

We acknowledge B.~Reichert for the participation in the initial stage of this project.

\textit{Note added.} During the preparation of this paper, a related preprint that contains a study of the mean-field contribution to the induced interaction appeared \cite{will_polaron_2021}. It reports an expression for the interaction that differs from our Eq.~(\ref{eq:E0}) by the $r$-independent term $2E^{\B}(\widetilde{G})$.


%

\end{document}